\documentclass[twocolumn,floats,prl, superscriptaddress,floatfix]{revtex4-2} 
\usepackage{times} 
\usepackage{amssymb,amsmath}
\usepackage{csquotes}
\usepackage{physics}
\usepackage{bm}
\usepackage{nicefrac}
\usepackage{mathtools}
\usepackage{setspace}
\usepackage{makecell}

\usepackage[usenames,dvipsnames]{color}
\usepackage[pdftex]{epstopdf}
\usepackage[pdftex]{graphicx}
\usepackage{hyperref}
\hypersetup{pdftitle={From bipedal to chaotic motion of chemically fueled partially wetting liquid drops} 
    pdfauthor={Florian Voss and Uwe Thiele},  
pdfproducer={lateX},
pdfview=FitV,       
pdfstartview=FitB,
linkcolor=blue,     
citecolor=blue,     
urlcolor=blue,      
breaklinks=true,    
colorlinks=true,
citebordercolor=0 0 0,  
filebordercolor=0 0 0,
linkbordercolor=0 0 0,
menubordercolor=0 0 0,
urlbordercolor=0 0 0,
pdfhighlight=/I,
pdfborder=0 0 0,   
bookmarksopen=true,
bookmarksnumbered=true
}
\DeclareGraphicsExtensions{.jpg, .pdf, .tif, .png}
\usepackage[usenames,dvipsnames]{color}
\usepackage[normalem]{ulem}
\newcommand{\subref}[2]{\hyperref[#1]{\ref{#1}#2}}
\newcommand{\vecg}[1]{\boldsymbol{#1}}

\newcommand{\tensg}[1]{\underline{\boldsymbol{#1}}}

\usepackage{booktabs}
\usepackage{multirow}
\makeatletter
\renewcommand{\p@subsection}{}
\renewcommand{\p@subsubsection}{}
\renewcommand{\@seccntformat}[1]{\csname the#1\endcsname\quad}
\makeatother

\begin{document}
\count\footins = 1000 

\title{From bipedal to chaotic motion of chemically fueled partially wetting liquid drops}

\author{Florian Voss}
\email{f\_voss09@uni-muenster.de}
\thanks{ORCID ID: 0009-0003-9679-035X}
\affiliation{Institute of Theoretical Physics, University of M\"unster, Wilhelm-Klemm-Str.\ 9, 48149 M\"unster, Germany}

\author{Uwe Thiele}
\email{u.thiele@uni-muenster.de}
\homepage{http://www.uwethiele.de}
\thanks{ORCID ID: 0000-0001-7989-9271}
\affiliation{Institute of Theoretical Physics, University of M\"unster, Wilhelm-Klemm-Str.\ 9, 48149 M\"unster, Germany}
\affiliation{Center for Data Science and Complexity (CDSC), University of M\"unster, Corrensstr.\ 2, 48149 M\"unster, Germany}

\begin{abstract}
  We employ a thermodynamically consistent out-of-equilibrium continuum model to study the motion patterns of partially wetting liquid drops covered by autocatalytically reacting surfactants. When ambient chemostats feed a chemomechanical feedback loop involving a nonlinear reaction network, surface stresses caused by the Marangoni effect and the ensuing hydrodynamic motion, drops show a variety of increasingly complex biomimetic motility modes including shuttling, bipedal, rotational, intermittently chaotic and chaotic motion. We determine the corresponding nonequilibrium phase diagram and show that the complexity of the motion arises from competing length scales.
\end{abstract}

\maketitle
The coupling of chemical and mechanical processes lies at the core of many intriguing phenomena in soft matter systems that are kept away from thermodynamic equilibrium. In living systems, such chemomechanical coupling is essential for, e.g., the division, differentiation and motility of individual cells \citep{WTAB2010n,GrKG2017arb,ChHH2017cb,SKLR2010scdb,DuBE2025pre,DBAP2025prl}, and for the spreading, dewetting and migration of cell aggregates and monolayers \cite{DGND2011pnasusa,BSKC2014pnasusa,BBDD2018pnasusa,PABG2019np,FCWD2025pnasusa}. This includes transitions between persistent, bipedal and random modes of migration \cite{BBDD2018pnasusa}. Such complex dynamic phenomena are often associated with intricate feedback loops between chemical signaling and active mechanical stresses.

Already in liquids such couplings naturally occur in the presence of interfaces where surface-active chemical species (surfactants) accumulate, thereby generically reducing the interfacial tension \citep{OrDB1997rmp,CrMa2009rmp,ThAP2016prf}. This then implies that corresponding concentration gradients directly result in mechanical stresses that drive so-called Marangoni flows, ultimately allowing for chemomechanical feedback loops. In consequence, chemical reactions and similar processes like solubilization, adsorption, dissolution and evaporation can lead to various forms of droplet (microswimmer) motion \citep{Mich2023arfm,SeFM2016epjt,KAMY2002jcp}, facilitate droplet division \citep{NSKY2005pre,BWBG2010ac} and may even result in persistent highly complex spatiotemporal patterns \citep{EATP2012c,PBKB2011acie,SYM2023sr}.

Cases of chemomechanical coupling where partially wetting drops move on a solid or liquid substrate are particularly interesting \cite{DoOn1995prl,SMHY2005prl,NSKY2005pre,ShBM2010e,PiAn2014cocis,SuYo2014epjt,SSK2017sm} as they can serve as simple biomimetic models for crawling cells and migrating aggregates \cite{Hanc2011ptrsbs}. In this way, one is able to determine minimal ingredients of a physico-chemical system needed to obtain complex motility modes. Accordingly, one may develop theoretical descriptions of such chemomechanical out-of-equilibrium systems that build on well-known thermodynamically consistent dynamic models for sessile drops \cite{Thie2018csa,ThAP2016prf,HDGT2024l}.
Note that existing continuum descriptions typically focus either on the chemical \citep{BMKY2013pre,YEMC2015sr} or the mechanical aspect \citep{DoKr2011prl,RAEB2013prl,KoPi2013sm}, or take a phenomenological phase-field approach \citep{ZiSA2012jrsi,DrAK2014njp}. In general, it is not guaranteed that the corresponding simplifications preserve thermodynamic consistency,~i.e., no physical limiting case exists where thermodynamic constraints \citep{RaEs2016prx,APFE2021jcp,AvAFE2024jcp,JuGS2018rpp} are satisfied. Preserving the thermodynamic core within an out-of-equilibrium model, e.g., by adding the exchange with external reservoirs, ensures that physically relevant reciprocal couplings are accounted for \citep{Onsa1931pr,Zwic2022cocis,VoTh2024jem}. Besides facilitating the quantification of dissipation and energetic costs for nonequilibrium states \citep{MFTC2021prx,AvAFE2024jcp,BiEs2025nc,PFME2025pre}, even novel physical effects may then be derived from these cross-couplings \citep{ThTL2013prl,GoFr2019prl}.

\begin{figure}
\centering
\includegraphics[scale=0.55]{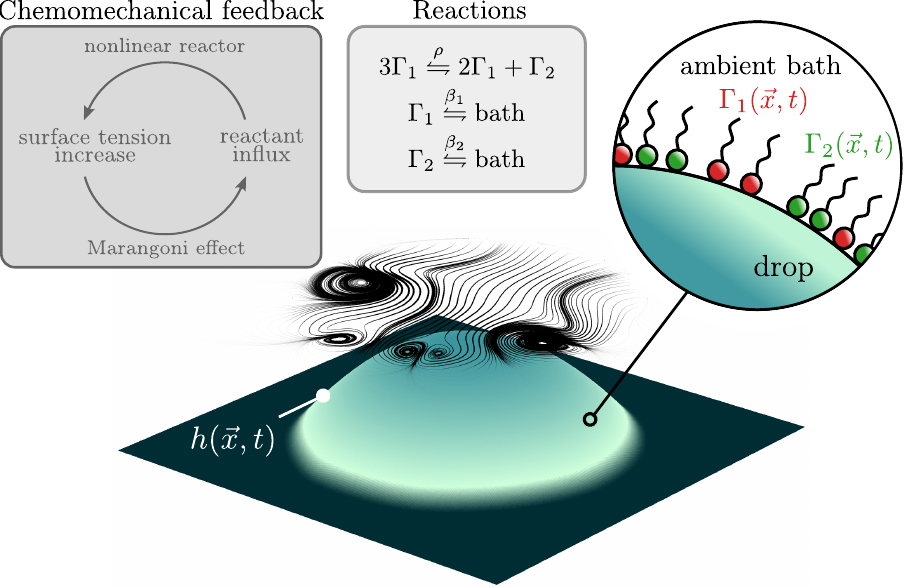}
\caption{Schematics of the considered system: drops of a partially wetting liquid on a flat solid substrate are described by a height profile $h(\vec{x}, t)$. The free surface is covered by two species of surfactants with density profiles $\Gamma_1(\vec{x},t)$ and $\Gamma_2(\vec{x}, t)$. They engage in an autocatalytic reaction and are exchanged with ambient reservoirs. The open exchange of fuel and waste together with a chemomechanical feedback loop drives droplet motility. The streamlines represent the height-integrated liquid flow.}
\label{fig:system_sketch}
\end{figure}

\begin{figure}
\centering
\includegraphics[width=0.49\textwidth]{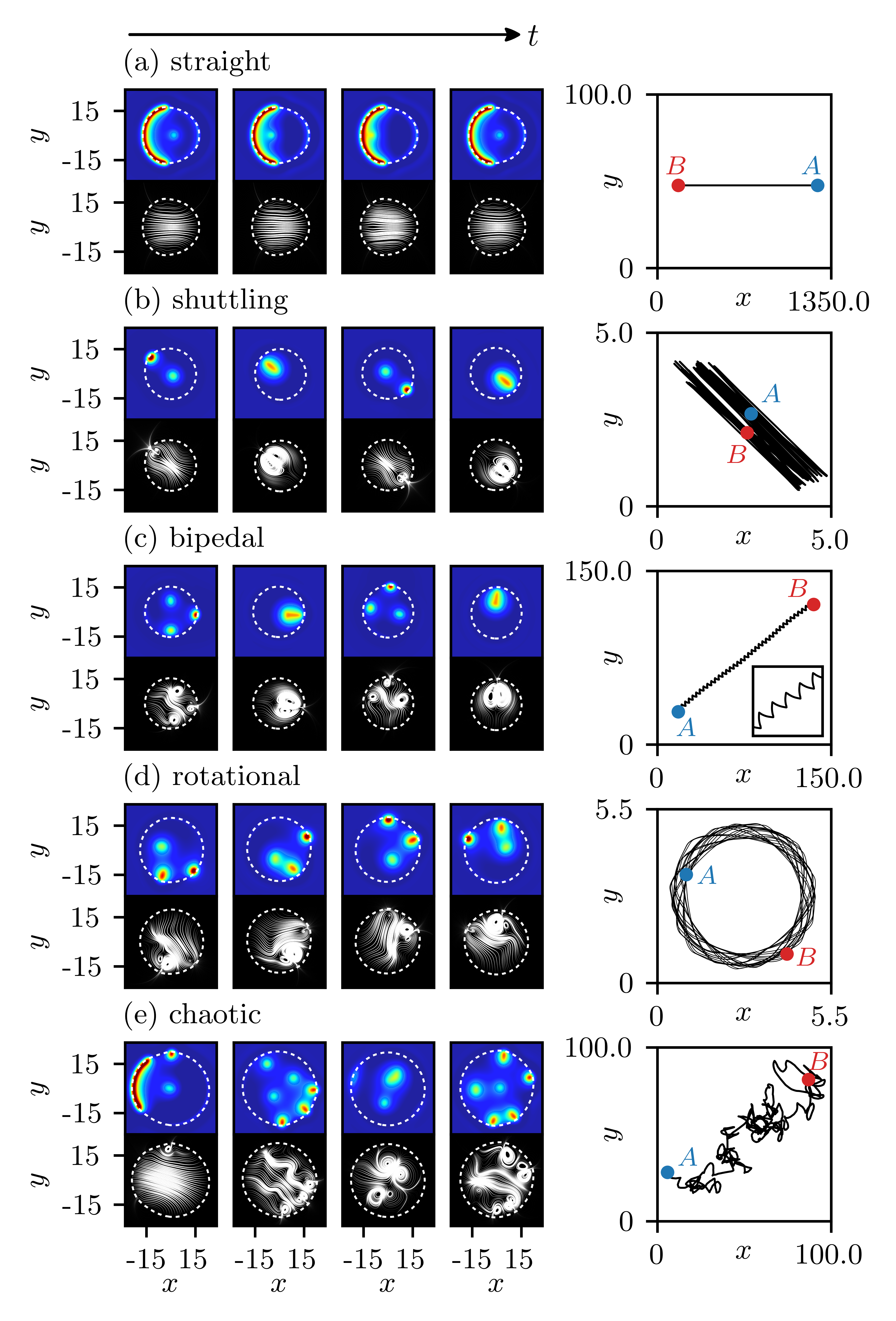}
\caption{Panels (a) to (e) give various motility modes of chemically fueled drops covered by reacting surfactants. Here, each top [bottom] row of the left part shows snapshots of surface tension profiles where red (blue) indicates high (low) values [the height-integrated liquid flow, namely, the flux of field $h$ in Eq.~(\ref{eq:gradient_dynamics_general})]. The streamline thickness represents the magnitude of the flux, and the dashed drop contours correspond to $h=1.1$  ($h=1.1h_a$ in dimensional units). The right parts give associated center-of-mass trajectories of the drop between points $A$ and $B$ on the substrate. In (c), the inset shows a magnification of the trajectory. Only part of the periodic computational domain $[-50,50]\times[-50,50]$ is shown. For parameters, see Appendix A. Also see Supplemental Videos 1-5.}
\label{fig:state_collection}
\end{figure}

Here, we employ a fully reciprocal, i.e., thermodynamically consistent model for a biomimetic open system with chemomechanical coupling and show how a variety of motility modes emerge. Specifically, we consider sessile drops of partially wetting liquids covered by two species of autocatalytically reacting insoluble surfactants that are exchanged with an ambient bath (Fig.~\ref{fig:system_sketch}). The bath acts as a chemostat continuously providing the drop with chemical fuel and removing waste. When using the chemostats to drive the droplets out of equilibrium, a rich spectrum of self-organized motility modes emerges: drops not only self-propel along straight lines but can, e.g., rotate, shuttle back and forth, perform directed bipedal motion and even show chaotic motion (Fig.~\ref{fig:state_collection}). As further explained below, all these complex motility modes result from a chemomechanical feedback loop between the nonlinear chemical reaction network on the drop surface and the solutal Marangoni effect. Importantly, they are not at all related to patterns already formed by the isolated surfactant system. In fact, the reaction-diffusion (RD) subsystem comprising the surfactants simply relaxes to a uniform steady state when decoupled from the drop \citep{VoTh2025prf}. This contrasts common descriptions of, e.g., the motion of Belousov-Zhabotinsky drops \citep{KAMY2002jcp,KYNS2011pre}, self-oscillating biomimetic gels \citep{YaBa2006s,LeDS2020prl} and motile cells on substrates \citep{DoKr2011prl,DrAK2014njp}, where the dynamics is driven by waves generated by the chemical subsystem. It also differs from descriptions of drops of active liquids involving polarization fields where microscopic self-propulsion and/or active stresses are assumed to exist without explicit incorporation of the underlying chemomechanical coupling \cite{JoRa2012jfm,KhAl2015pre,TjMC2012pnasusa,TTMC2015nc,LoEL2020sm,StJT2022sm}. 


To describe the motion of 3D drops on 2D substrates we employ the mesoscopic hydrodynamic model \citep{VoTh2025prf}
\begin{equation}
\partial_t \vecg{\psi} \left(\vec{x},t\right) = \vec{\nabla}\cdot\left[\tensg{\mathcal{M}}\left(\vecg{\psi}\right)\vec{\nabla}\frac{\delta\mathcal{F}}{\delta\vecg{\psi}}\right]+\vecg{\mathcal{R}}\left(\vecg{\psi}\right), \label{eq:gradient_dynamics_general}
\end{equation}
where $\vecg{\psi} = (h, \Gamma_1, \Gamma_2)^\text{T}$ represents the dynamic fields, namely, the drop height profile $h$ and the two surfactant density profiles $\Gamma_1, \Gamma_2$ (particles per unit interface area), all parametrized in substrate coordinates $\vec{x}=(x,y)^\text{T}$. We assume shallow drops, i.e., we employ a long-wave approximation \citep{CrMa2009rmp,BEIM2009rmp}. The corresponding free energy $\mathcal{F}=\int[f(h)+g\left(\Gamma_1,\Gamma_2\right)+\frac{\gamma_0}{2}\vert\vec{\nabla}h\vert^2]\:\text{d}^2x$ comprises three contributions. The wetting energy $f(h)=A(-\frac{1}{2h^2}+\frac{h_a^3}{5h^5})$, with the Hamaker constant $A>0$, encodes partial wetting. The minimum at $h=h_a$ corresponds to the thickness of the adsorption layer that coexists with the drop at the mesoscopic level. Assuming a sparse drop coverage by surfactants, their free energy is purely entropic, $g(\Gamma_1, \Gamma_2) = \gamma_0+k_{\rm B}T\sum_{i=1}^2\Gamma_i[\ln(\Gamma_ia^2_i)-1]$. Here, $\gamma_0$ is the surface tension of the bare drop, $k_{\rm B}T$ is the thermal energy, and $a_1, a_2$ are characteristic surfactant length scales. This choice results in the linear equation of state $\gamma(\Gamma_1, \Gamma_2) = \gamma_0 -k_{\rm B}T(\Gamma_1+\Gamma_2)$ \citep{ThAP2012pf, ThAP2016prf},~i.e., the local surface tension $\gamma$ depends only on the total surfactant density. The final term in $\mathcal{F}$ penalizes gradients in $h$. The variational derivatives $\delta\mathcal{F}/\delta h, \delta\mathcal{F}/\delta \Gamma_1$ and $\delta\mathcal{F}/\delta \Gamma_2$ are the liquid pressure and the surfactant chemical potentials on the drop surface, respectively. In contrast to recent models for chemically active bulk drops \citep{KiZw2021jotrsi,Zwic2022cocis,DGMF2023prl,GDMF2024prr,RBZM2025},  the positive definite symmetric mobility matrix $\tensg{\mathcal{M}}$ is fully occupied with nonlinear coefficients, 
 \begin{equation}
	\tensg{\mathcal{M}} = \left(\begin{array}{ccc}
	\frac{h^3}{3\eta} & \frac{h^2\Gamma_1}{2\eta} & \frac{h^2\Gamma_2}{2\eta} \\ 
	\frac{h^2\Gamma_1}{2\eta} & \frac{h\Gamma_1^2}{\eta}+D_1\Gamma_1 & \frac{h\Gamma_1\Gamma_2}{\eta}\\ 
	\frac{h^2\Gamma_2}{2\eta} &  \frac{h\Gamma_1\Gamma_2}{\eta}  &  \frac{h\Gamma_2^2}{\eta}+D_2\Gamma_2
	\end{array} \right),\label{eq:mobility_M}
\end{equation}
where $\eta$ is the dynamic viscosity of the liquid and $D_1, D_2$ are surfactant diffusivities. To exclude spatial pattern-forming instabilities of the decoupled RD subsystem, here, we assume equal diffusivities $D_1=D_2=D$. Eq.~(\ref{eq:mobility_M}) can be obtained from a long-wave expansion of the associated Stokes problem (cf.\ Refs.~\citep{PTTK2007pf, ThAP2016prf}). Note that the nonlinear entries in (\ref{eq:mobility_M}) \enquote{translate} the driving gradients of liquid pressure \textit{and} chemical potential of surfactant into hydrodynamic transport,~i.e., they encode the nonlinear mechanical responses to the driving gradients \citep{Note1}.
The surfactants undergo the reversible autocatalysis $3\Gamma_1 \stackrel{\rho}{\leftrightharpoons}2\Gamma_1+\Gamma_2$ with the rate constant $\rho>0$. This choice represents a central nonlinearity in several paradigmatic models for chemical pattern formation \citep{PrLe1968jcp,Nicolis1999}. Specifically for surfactants, such higher-order reactions may describe the formation of mixed micelles \citep{SSK2017sm,TaNN2021sm}. For the reversible exchange with the reservoirs $\Gamma_1 \stackrel{\beta_1}{\leftrightharpoons} \text{reservoir } (\mu_1^\mathrm{ex}),
\Gamma_2 \stackrel{\beta_2}{\leftrightharpoons} \text{reservoir }(\mu_2^\mathrm{ex})$ with the rate constants $\beta_1, \beta_2>0$, we assume uniform and constant external chemical potentials $\mu_1^\mathrm{ex}, \mu_2^\mathrm{ex}$. The system is driven out of equilibrium precisely when $\mu_1^\mathrm{ex}\neq\mu_2^\mathrm{ex}$ \citep{AvAFE2024jcp, VoTh2025prf}. The total reactive current in (\ref{eq:gradient_dynamics_general}) is $\vecg{\mathcal{R}} = (0,\mathcal{A}+\mathcal{B}_1,-\mathcal{A}+\mathcal{B}_2)^\text{T}$, where $\mathcal{A}$ and $\mathcal{B}_1,\mathcal{B}_2$ are the currents of the autocatalysis and of the reservoir exchanges, respectively. They are given by $\mathcal{A} = \rho\left[\exp\left(\frac{2}{k_{\rm B} T}\frac{\delta F}{\delta \Gamma_1}+\frac{1}{k_{\rm B} T}\frac{\delta F}{\delta \Gamma_2}\right)-\exp\left(\frac{3}{k_{\rm B} T}\frac{\delta F}{\delta \Gamma_1}\right)\right]$ and $\mathcal{B}_{1,2}=\beta_{1,2}\left[\exp\left(\frac{\mu_{1,2}^\mathrm{ex}}{k_{\rm B} T}\right)-\exp\left(\frac{1}{k_{\rm B} T}\frac{\delta F}{\delta \Gamma_{1,2}}\right)\right]$. Importantly, they obey local detailed balance \citep{APFE2021jcp,AvAFE2024jcp}. Note that the liquid volume is conserved by Eq.~(\ref{eq:gradient_dynamics_general}). A nondimensionalization of the model (\ref{eq:gradient_dynamics_general})-(\ref{eq:mobility_M}) is given in Appendix D. In the following, all quantities are nondimensional.

Returning to the various modes of drop motility, we note that the dynamics is characterized by a continuous generation, motion and merging of reactive spots of increased surface tension that is accompanied by intricate flow patterns (Fig.~\ref{fig:state_collection} and Supplemental Videos 1-5). The only exception is the straight motion in Fig.~\subref{fig:state_collection}{(a)} that is induced by a crescent-shaped surface tension profile at the front \citep{Note2}. 
For shuttling motion [Fig.~\subref{fig:state_collection}{(b)}], a central spot periodically merges with another spot that subsequently appears on alternating sides of the drop, resulting in a back-and-forth shuttling. Bipedal and rotational motion [Fig.~\subref{fig:state_collection}{(c)} and~\subref{fig:state_collection}{(d)}] are possible when three spots are present. These modes respectively correspond to left-right alternating spot dynamics yielding a directed \enquote{zig-zag} motion, and to discrete unidirectional turns of the spot pattern around the drop center in between merges. If more spots are present, the dynamics becomes chaotic and drops move irregularly across the substrate [Fig.~\subref{fig:state_collection}{(e)}]. In this chaotic regime at large times, drop motion is approximately diffusive \citep{VoTh2025prf},~i.e., it is (quasi-)random. However, because the transition to chaos occurs via intermittency, drops can also perform chaotic motion interspersed with bipedal or rotational motion (Appendix~E), reminiscent of L\'evy flights \citep{MeKl2004jpa} and run-and-tumble motion \citep{CaTa2015arcmp}, respectively.

An inspection of the drop center-of-mass trajectories (Fig.~\ref{fig:state_collection}) reveals that all motility modes (except straight motion) are in fact somewhat irregular, i.e., periodic motion (bipedal motion, shuttling) is superimposed with some drift or the motion is quasiperiodic (rotational motion). This indicates that these states are nested deeply in the underlying highly complicated bifurcation structure. 
The foregoing observations leave us with two central questions. First, since this striking dynamics does not originate from pattern formation already present in the underlying RD subsystem, what causes the continuous generation and merging of spots? And second, can we identify quantities that determine the emerging motility mode, even in the studied strongly nonlinear regime?

The origin of the observed complex dynamics is a chemomechanical feedback loop between the reaction network on the drop surface and the Marangoni effect (Fig.~\ref{fig:system_sketch}). We outline this interplay here and refer to Ref.~\citep{VoTh2025prf} for details. The positive feedback stems from the capacity of the reaction network to increase the local surface tension in response to an \textit{influx} of surfactants from neighboring regions on the drop. This circumstance is rather counterintuitive since an influx of surfactants is expected to \textit{increase} the local overall surfactant mass, thereby \textit{reducing} the surface tension. Here, the opposite occurs due to the nonlinearity of the (open) reaction network, which locally allows for a net removal of surfactant from the drop when the influx is sufficiently large. This may then raise the local surface tension, which in turn further increases the influx due to the Marangoni effect. This closes the feedback loop and leads to a fast generation of spots of increased surface tension in response to sufficiently large flow perturbations. Due to its high surface tension, each spot is a center of the total surfactant flow. As a result, multiple spots are always transported towards each other, i.e., the observed spot merging is directly caused by the Marangoni stresses induced by individual spots. The change in the surfactant flow resulting from the merges again excites the generation of new spots elsewhere on the drop, thereby leading to the observed perpetual dynamics.

\begin{figure}
\centering
\includegraphics[scale=0.68]{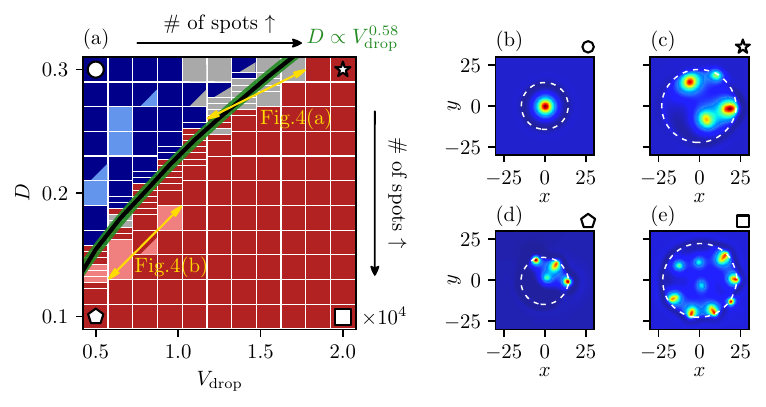}
\caption{(a) Morphological phase diagram spanned by surfactant diffusivity $D$ and drop volume $V_\text{drop}$. States are represented by colors: resting drop (dark blue), shuttling (light blue), rotational (gray), bipedal (light red) and chaotic (dark red) motion. The number of spots increases with decreasing $D$ and increasing $V_\text{drop}$. The primary instability of the resting drop state is determined by continuation (black line) and fitted with the power law $D\propto V^{0.58}_\text{drop}$ (green line). Colored triangular corners indicate multistability with the respective state. Yellow lines indicate slices analyzed in Fig.~\ref{fig:parameter_slices}. (b)-(e) Snapshots of the surface tension profiles corresponding to the four diagram corners (style as in Fig.~\ref{fig:state_collection}). For parameters, see Appendix A.}
\label{fig:phase_diagram}
\end{figure}

To understand what determines the arising motility modes, we first note that the states shown in Figs.~\subref{fig:state_collection}{(b)-(e)} indicate a positive correlation between the maximal occurring spot number and the complexity of the state. A natural assumption is that the spot number is related to the drop size and the characteristic spot size. That is, larger drops and smaller spots allow for a larger number of spots. We test this hypothesis by performing a parameter scan (Appendix~D) in the surfactant diffusivity $D$ and the droplet volume $V_\text{drop} = \int [h(\vec{x},t)-h_a]\:\text{d}^2x$ that control the spot and drop size, respectively [Fig.~\subref{fig:phase_diagram}{(a)}]. Initiating each time simulation by a flat film and noisy surfactant densities, drops form by spinodal dewetting and normally coarsen into a single drop. Reducing $D$ or increasing $V_\text{drop}$ indeed allows drops to accommodate more spots. In particular, small $D$ and large $V_\text{drop}$ result in many small spots and highly irregular drop motion [Fig.~\subref{fig:phase_diagram}{(e)}]. In the opposite case (large $D$, small $V_\text{drop}$), drops are at rest in a nonequilibrium steady state, characterized by a single large spot at the drop center [Fig.~\subref{fig:phase_diagram}{(b)}]. The other corners of the phase diagram [Figs.~\subref{fig:phase_diagram}{(c)} and \subref{fig:phase_diagram}{(d)}] also correspond to chaotic motion, but with fewer spots. Shuttling, rotational and bipedal motion all occur at intermediate values of $D$ and $V_\text{drop}$. We remark that straight motion as shown in Fig.~\subref{fig:state_collection}{(a)} does not emerge from the dewetting of a flat film and therefore does not appear in Fig.~\subref{fig:phase_diagram}{(a)}.

\begin{figure}
\centering
\includegraphics[scale=0.85]{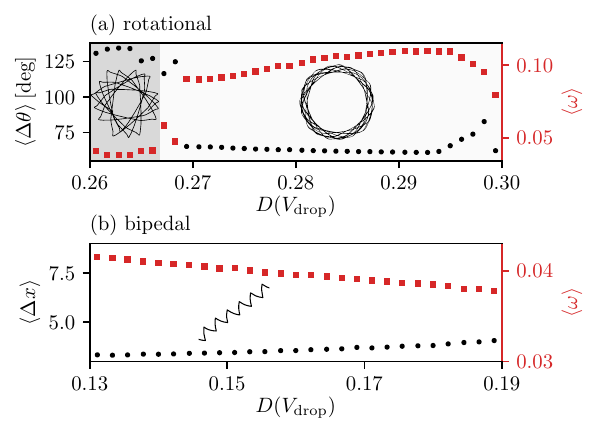}
\caption{One-parameter diagram along the lines marked in Fig.~\subref{fig:phase_diagram}{(a)}, i.e., the diffusivity $D(V_\text{drop})$ is linearly related to $V_\text{drop}$. Shown are (a) the turning angle $\langle\Delta\theta\rangle$ and (b) the step size $\langle\Delta x\rangle$, along with the deformation frequency $\langle\omega \rangle$. For drop rotation, there is a transition between modes with star-shaped (gray region) and circular trajectories (white region). Trajectories are indicative of the respective mode.} 
\label{fig:parameter_slices}
\end{figure}

We further quantify rotational and bipedal motion by a time-averaged characteristic frequency $\langle \omega\rangle$, corresponding to their turn and step frequency, respectively (Appendix B). We also compute the time-averaged turning angle $\langle \Delta \theta\rangle$ and step size $\langle \Delta x \rangle$. Fig.~\ref{fig:parameter_slices} shows these quantities along the lines marked in Fig.~\subref{fig:phase_diagram}{(a)}. For rotational motion, we find two different modes characterized by star-shaped and circular trajectories of the drop's center-of-mass. The transition between them appears as a jump in $\langle \omega\rangle$ and $\langle \Delta\theta\rangle$. In the star-shaped mode, drops undergo turns at a lower frequency but perform larger rotations, while the opposite is true for the circular mode. Near the transition, we also find circular modes with similar $\langle \omega\rangle$ and $\langle \Delta\theta\rangle$ as the star-shaped one. Note that both modes furthermore allow for superimposed drift. For the circular mode, this occurs at larger $D$ and coincides with a decrease of $\langle \omega \rangle$  [increase of $\langle \Delta\theta \rangle$], because in the presence of drift our definitions of $\langle \Delta \theta\rangle$ and $\langle \Delta x \rangle$ no longer reflect the turn frequency and angle (see Appendix B). For bipedal motion, the step frequency [size] continuously decreases [increases] with increasing $D$.

Finally, we return to the overall transition between resting and irregularly moving drops in Fig.~\subref{fig:phase_diagram}{(a)}. Employing numerical continuation for the resting drop, we identify the threshold of instability. Tracing it in the $(V_\text{drop}, D)$-plane gives the black line in Fig.~\subref{fig:phase_diagram}{(a)} that well captures the overall transition. However, the corresponding bifurcation is a subcritical pitchfork bifurcation (Appendix C) indicating that resting drops may be multistable with other states. Performing additional time simulations initiated with states obtained at nearby points in the $(D,V_\text{drop})$-plane, we confirm multistability with shuttling, rotating and chaotically moving drops [Fig.~\subref{fig:phase_diagram}{(a)}]. Further, employing a scaling argument, we reconnect this instability to the observation that the drop dynamics is determined by a competition of length scales of the drop and spot sizes. To this end, we introduce the length scale $L_\text{drop}\propto V_\text{drop}^{1/3}$, which represents the radius of the drop footprint. We further assume that the characteristic spot size is determined as a diffusion length $L_\text{spot}\propto D^{1/2}$. Assuming that the instability occurs when the ratio $L_\text{drop}/L_\text{spot}$ crosses a certain threshold, we obtain the scaling $D\propto V_\text{drop}^{2/3}$. Fitting the instability curve in Fig.~\subref{fig:phase_diagram}{(a)} as $D(V_\text{drop}) = \alpha V_\text{drop}^\beta$ yields $\beta\approx 0.58$ [green line in Fig.~\subref{fig:phase_diagram}{(a)}]. Despite its simplicity, the given scaling argument qualitatively captures the instability threshold \citep{Note3}. That is, destabilization of the resting drop occurs \textit{approximately} at a critical ratio of drop and spot size.

To conclude, we have employed a thermodynamically consistent out-of-equilibrium model to study the various biomimetic modes of motility of partially wetting liquid drops covered by autocatalytically reacting surfactants. Specifically, ambient chemostats drive the drops out of equilibrium, thus sustaining a chemomechanical feedback loop involving a nonlinear reaction network and the Marangoni effect. Here, this results in increasingly complex motility patterns involving shuttling, bipedal, rotational and irregular motion, with complexity arising from competing length scales. We expect that similar chemomechanical feedback loops are relevant to various biomimetic drop systems with nonlinearly reacting surfactants \citep{SSK2017sm,TaNN2021sm,SuYo2014epjt,KAMY2002jcp,KYNS2011pre}. The higher-order chemical reactions may, e.g., represent the formation of mixed surfactant micelles. More generally, we have demonstrated that complex motility modes that are also found in genuine biological contexts \citep{BBDD2018pnasusa,LZGA2015jrsi,KLBE2019po,CZLL2013prl,DrAK2014njp,LoZA2014sm,BAJT2010bj} can emerge and be analyzed in relatively simple nonequilibrium settings where the chemomechanical interplay can be treated in a transparent  thermodynamically consistent manner. We emphasize that the underlying thermodynamic structure is generic and further found in recently studied models for chemically active mixtures \citep{KiZw2021jotrsi,Zwic2022cocis,DGMF2023prl,GDMF2024prr} and various other soft matter systems \citep{Doi2011jpcm}. Therefore, these systems may display similar phenomena when higher-order chemical reactions and nonlinear, fully occupied mobilities (here, encoding hydrodynamics) are considered. The latter may alternatively result from (cross-)diffusion in crowded environments \citep{FaMc2010pre} or from mechanical effects related to elasticity.

\textit{Acknowledgements---}Part of the calculations for this publication were performed on the HPC cluster PALMA II
of the University of Münster, subsidized by the Deutsche Forschungsgemeinschaft (DFG) (INST
211/667-1). We acknowledge financial support by the DPG via Grant No. TH 781/12-2 within SPP
2171. FV further acknowledges valuable discussions with Yutaka Sumino. 

\textit{Data availability---}The data and python codes used in this study as well as the supplemental videos referenced in the main text will be made publicly available at \citep{VoTh2025_26zn}.

\appendix
\section{End Matter}
\textit{Appendix A: Nondimensionalization and parameter sets---}We rescale Eqs.~(\ref{eq:gradient_dynamics_general}) by introducing the scaling $t = \tau\tilde{t},  (x,y)=L(\tilde{x}, \tilde{y}), h=l\tilde{h} ,\Gamma_\alpha=\tilde{\Gamma}_\alpha/(a_1a_2), (f, g) = \kappa(\tilde{f}, \tilde{g})$, where $\alpha=1,2$ and dimensionless quantities are denoted by a tilde. The scales are chosen as $\tau = \frac{L^2\eta}{\kappa l}, L=\sqrt{\frac{\gamma_0}{\kappa}}l, l=h_a,  \kappa=\frac{k_{\rm B} T}{a_1 a_2}$. This yields the nondimensional parameters
\begin{align}
\begin{split}  
\delta &= \frac{a_1}{a_2}, \qquad
W = \frac{A}{l^2\kappa}, \qquad \tilde{D}_\alpha = \frac{\tau}{L^2}k_{\rm B}T D_\alpha, \\
\tilde{r} &= \tau a_1a_2 r, \qquad
\tilde{\beta}_\alpha = \tau a_1a_2 \beta_\alpha, \qquad
\tilde{\mu}^\mathrm{ex}_\alpha = \frac{1}{k_{\rm B}T}\mu^\mathrm{ex}_\alpha.
\end{split}
\label{app:eq:nondimensional_parameters}
\end{align} 
Omitting tildes, the dimensionless equations are
\begin{align}
\begin{split}
\partial_t h = &\vec{\nabla}\cdot\left[\frac{h^3}{3}\vec{\nabla} p +\frac{h^2}{2}\vec{\nabla}\left(\Gamma_1+\Gamma_2\right)\right],\\
\partial_t \Gamma_1 = &\vec{\nabla}\cdot\left[\frac{h^2\Gamma_1}{2}\vec{\nabla} +h\Gamma_1\vec{\nabla} \left(\Gamma_1+\Gamma_2\right)\right]\\
&+D_1\Delta\Gamma_1+\mathcal{A}+\mathcal{B}_1,\\
\partial_t \Gamma_2 = &\vec{\nabla}\cdot\left[\frac{h^2\Gamma_2}{2}\vec{\nabla} +h\Gamma_2\vec{\nabla}\left(\Gamma_1+\Gamma_2\right)\right]\\
&+D_2\Delta\Gamma_2-\mathcal{A}+\mathcal{B}_2,
\end{split}
\label{app:eq:model_nondim}
\end{align}
where the liquid pressure $p$ is given by
\begin{align}
p = W\left(\frac{1}{h^3}-\frac{1}{h^6}\right)-\Delta h,\label{app:eq:pressure_nondim}
\end{align}
and the reaction terms are
\begin{align}
\begin{split}
\mathcal{A} &= r\left[\delta\Gamma_2\Gamma_1^2-(\delta\Gamma_1)^3\right], \\
\mathcal{B}_1&=\beta_1\left[e^{\mu_1^\mathrm{ex}}-\delta\Gamma_1\right], \\
\mathcal{B}_2&=\beta_2\left[e^{\mu_2^\mathrm{ex}}-\delta^{-1}\Gamma_2\right].
\end{split}
\end{align}
We list the parameter sets for all figures in Table~\ref{app:tab:parameter_sets}.

\begin{table*} 
\caption{List of parameter sets for each figure.}
\begin{tabular}{cccccc}
\toprule\midrule
Figure & Domain & Numerical grid & $D$& $V_\text{drop}$ & Other parameters \\  \midrule
\subref{fig:state_collection}{(a)} & $[-50,50]\times[-50,50]$& $251\times 251$ & 0.12 & $0.80\times 10^4$ & \makecell{$W=10$, $\rho = 0.3$, $\beta_1=2$,$\beta_2=0.01$, \\$\mu^\mathrm{ex}_1=-1.4$, $\mu^\mathrm{ex}_2=4.175$, $\delta=1$}\\
\subref{fig:state_collection}{(b)} & as in \subref{fig:state_collection}{(a)} & as in \subref{fig:state_collection}{(a)}& 0.24 & $0.65\times 10^4$ & as in \subref{fig:state_collection}{(a)} \\
\subref{fig:state_collection}{(c)} & as in \subref{fig:state_collection}{(a)} & as in \subref{fig:state_collection}{(a)}& 0.14 & $0.65\times 10^4$ & as in \subref{fig:state_collection}{(a)} \\
\subref{fig:state_collection}{(d)} & as in \subref{fig:state_collection}{(a)} & as in \subref{fig:state_collection}{(a)}& 0.2683 & $1.30\times 10^4$ & as in \subref{fig:state_collection}{(a)} \\
\subref{fig:state_collection}{(e)} & as in \subref{fig:state_collection}{(a)} & as in \subref{fig:state_collection}{(a)}& 0.2 & $2.00\times 10^4$ & as in \subref{fig:state_collection}{(a)} \\
\ref{fig:phase_diagram} & \makecell{$[-50,50]\times[-50,50]$ (time sim.)\\ varying (continuation)}& \makecell{$251\times 251$ (time sim.)\\ varying (continuation)}& varying & varying & as in \subref{fig:state_collection}{(a)} \\
\ref{fig:bifurcation_slice} & $[-50,50]\times[-50,50]$ & $120\times 120$ & varying & $0.95\times10^4$  & as in \subref{fig:state_collection}{(a)} \\
\subref{fig:motion_w_intermittency}{(a)} & as in \subref{fig:state_collection}{(a)} & as in \subref{fig:state_collection}{(a)} & 0.145 &  $0.50\times10^4$  & as in \subref{fig:state_collection}{(a)} \\
\subref{fig:motion_w_intermittency}{(b)} &as in \subref{fig:state_collection}{(a)} & as in \subref{fig:state_collection}{(a)} & 0.1267 & $0.55\times10^4$  & as in \subref{fig:state_collection}{(a)} \\
\midrule
\bottomrule
\end{tabular} 
\label{app:tab:parameter_sets}
\end{table*}

\textit{Appendix B: Definitions of the characteristic quantities $\langle \omega \rangle, \langle \Delta \theta\rangle$ and $\langle \Delta x\rangle$---} As discussed in the main text, many of the obtained states are irregular, i.e., they feature complex temporal behavior that cannot be characterized by a single frequency. Nevertheless, for bipedal and rotational motion, a characteristic frequency can be defined by considering the time intervals between steps and turns, respectively. For bipedal motion, the corresponding frequency is straightforwardly obtained from the L$_2$-norm $\vert\vert h\vert\vert_2 (t) = \left(\int h(\vec{x},t)^2\:\text{d}^2x\right)^{1/2}$. Specifically, we define the time $T$ between consecutive maxima of $\vert\vert h\vert\vert_2 (t)$ and have $\langle \omega\rangle = \langle \frac{2\pi}{T}\rangle$, where $\langle \cdot\rangle$ denotes the temporal average over the trajectory. For rotational motion, we define instead the distance $R(t) = \left[(x_d-\langle x_d\rangle)^2+(y_d-\langle y_d\rangle)^2\right]^{1/2}$, where $x_d$ and $y_d$ are the coordinates of the drop's center-of-mass, given by $\vec{x}_d(t) = 1/V \int \vec{x} h(\vec{x},t)\:\text{d}^2x$. To sensibly handle the periodic boundaries, we use the algorithm presented in \citep{BaBre2008}. As before, the frequency $\langle \omega \rangle$ is then determined from the maxima of $R(t)$. Note that the definition of $R(t)$ employs the time-averaged center-of-mass as a center for the trajectory. Conceptually, this breaks down if rotation is superimposed with drift. Then, $\langle \omega \rangle$ no longer reflects the drop turn frequency. For bipedal motion, the mean step size $\langle\Delta x\rangle$ is defined as the average center-of-mass distance traveled between maxima of $\vert\vert h\vert\vert_2 (t)$. Similarly, the mean turn angle $\langle\Delta \theta\rangle$ is defined as the average angle between the vectors $(x_d-\langle x_d\rangle,y_d-\langle y_d\rangle)^T$ at consecutive maxima of $R(t)$.

\begin{figure}
\centering
\includegraphics[scale=0.67]{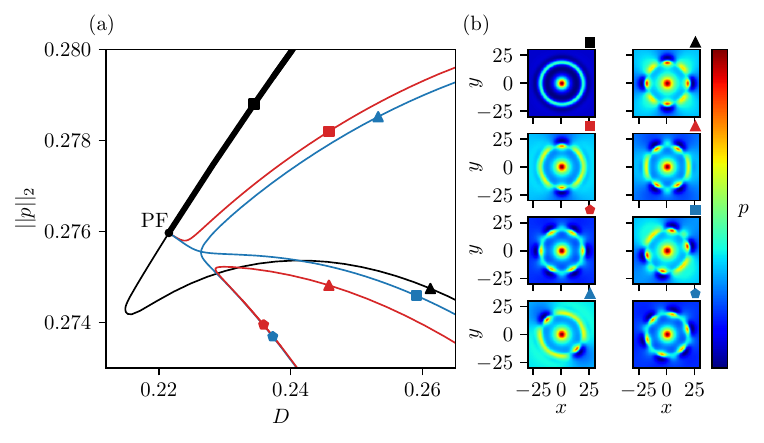}
\caption{(a) Partial bifurcation diagram of the originally stable resting drop (black line) with diffusivity $D$ as control parameter. The solution measure is the L$_2$-norm $\vert\vert p \vert\vert_2 = 1/L\left(\int  p^2\:\text{d}^2 x\right)^{1/2}$ where $p$ is given by Eq.~(\ref{app:eq:pressure_nondim}). The domain size is $100\times100$. Strong [weak] lines represent linearly stable [unstable] states. The stable resting drop is destabilized in a subcritical pitchfork bifurcation (PF). Note that other occurring bifurcations are not marked. Red and blue lines indicate branches of unstable states that bifurcate at the PF. The disconnected branches of the same color are related by an imperfect pitchfork bifurcation. (b) Liquid pressure profiles of the different emerging states [marked by corresponding symbols in (a)]. Red (blue) indicates high (low) pressure. For parameters, see Table~\ref{app:tab:parameter_sets}.}
\label{fig:bifurcation_slice}
\end{figure}

\textit{Appendix C: Partial bifurcation diagram of the resting drop---}Fig.~\subref{fig:bifurcation_slice}{(a)} shows a partial bifurcation diagram of the resting drop states, employing $D$ as control parameter. It corresponds to a vertical crossing of the instability curve in Fig.~\subref{fig:phase_diagram}{(a)} at $V_\text{drop} = 0.95\times 10^4$. The stable resting drop is not exactly radial but due to the quadratic domain only exactly invariant under rotation by $\pi/2$. It is destabilized at a subcritical pitchfork bifurcation (PF) of multiplicity two, i.e., two branches of steady states emerge (red and blue). They correspond to drops invariant w.r.t.\ reflection at the vertical and the diagonal, respectively, i.e., the states on all red and blue branches are approximately mutually related by $\pi/4$-rotation. Note that each of the lines themselves represent pairs of states related by $\pi/2$-rotation (the symmetry broken at PF). The structure of the states, in particular, their modulations in the contact-line region, is best visible in the pressure profiles shown in Fig.~\subref{fig:bifurcation_slice}{(b)}. Both branches (red and blue) emerging at the PF subsequently undergo an \textit{imperfect} PF, each characterized by an additional saddle-node bifurcation where a pair of new branches emerges. For the blue set of branches, this saddle-node bifurcation nearly coincides with the branch emerging at PF, whereas for the red set there is a noticeable gap. Note that the states on the branches marked by pentagon symbols nearly show a hexagonal symmetry. We expect that the saddle-node bifurcation visible for the black branch at low $D$ also corresponds to part of an imperfect PF. On a circular domain, the respective bifurcation breaks the rotational symmetry of the base state and a state with only $\pi/2$-rotational symmetry emerges. On a square domain, this already is the highest possible symmetry and hence the pitchfork is rendered imperfect. 

From Fig.~\ref{fig:bifurcation_slice}, we thus make two notable observations. First, the primary bifurcation of the stable resting drop is subcritical, i.e., one can expect a dramatic change in the dynamics once this threshold is crossed. Second, the emerging unstable states mostly feature modulations in the contact line region. Both of these observations are congruent with the results shown in Fig.~\ref{fig:phase_diagram} since crossing this instability threshold results in a sudden transition to various modes of drop motility that all feature prominent spot patterns in the contact line region.

\textit{Appendix D: Numerical details---}For direct numerical simulations, we discretize Eqs.~(\ref{app:eq:model_nondim})-(\ref{app:eq:pressure_nondim}) using the finite-element method implemented in the C++ library oomph-lib \citep{HeHa2006} and employ an adaptive backward differentiation scheme of order 2 for time stepping.
To obtain the one-parameter diagrams shown in Fig.~\ref{fig:parameter_slices}, we initialize time simulations from several states on the marked lines in the $(V_\text{drop},D)$-plane in Fig.~\subref{fig:phase_diagram}{(a)}, while slightly changing $V_\text{drop}$ and $D$ to obtain neighboring points and then iterate. For details on domain and grid, see Table~\ref{app:tab:parameter_sets}.

For numerical continuation, we use the finite element-based Matlab package pde2path \cite{Uecker2021}. Eqs.~(\ref{app:eq:model_nondim})-(\ref{app:eq:pressure_nondim}) are supplemented with auxiliary conditions resulting from spatial translation invariance and liquid volume conservation. Computations are again performed on a  periodic square domain. To compute the instability threshold in Fig.~\subref{fig:phase_diagram}{(a)}, we continue the resting drop state in $D$ at different fixed $V_\text{drop}$. We vary domain and grid size depending on the drop volume such that the adsorption layer only covers a small part of the domain, while approximately keeping the numerical resolution. Typical domain and grid sizes are $[-30,30]\times[-30,30]$ and  $145\times 145$. The bifurcation diagram in Fig.~\ref{fig:bifurcation_slice} is computed on the full $[-50,50]\times[-50,50]$ domain with a $120 \times 120$ grid.

\begin{figure}
\centering
\includegraphics[scale=0.9]{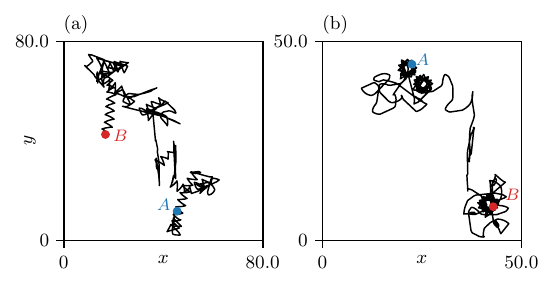}
\caption{Example trajectories of the drop's center-of-mass between points $A$ and $B$ in cases with intermittency. Drops show phases of chaotic motion interspersed with (a) bipedal motion and (b) rotational motion. Parameters are given in Table~\ref{app:tab:parameter_sets}.}
\label{fig:motion_w_intermittency}
\end{figure}

\textit{Appendix E: Intermittency---}Fig.~\ref{fig:motion_w_intermittency} shows trajectories of drops undergoing intermittent motion. In Fig.~\subref{fig:motion_w_intermittency}{(a)}, phases of chaotic motion are interspersed with bipedal directed motion. Fig.~\subref{fig:motion_w_intermittency}{(b)} is analogous but with interspersed rotational motion. In both cases, drops are small and the system is slightly beyond the instability threshold of the stable resting drop [cf. Table~\ref{app:tab:parameter_sets} and Fig.~\subref{fig:phase_diagram}{(a)}].

\end{document}